\documentclass[reprint,showpacs,prd,preprintnumbers,amssymb,nofootinbib,superscriptaddress,aps]{revtex4-1}
\usepackage{graphicx,floatflt,amssymb,rotate,epsfig}
\usepackage[utf8]{inputenc}
\usepackage[inline]{enumitem}
\usepackage{mathrsfs}
\usepackage{amssymb}
\usepackage{amsmath}
\usepackage{color}
\usepackage{graphicx}
\usepackage{subfig}
\usepackage{multirow}
\usepackage{mathtools}
\usepackage{float}
\usepackage{pstricks}
\usepackage[toc,page]{appendix}
\usepackage{pifont}
\usepackage{braket}
\usepackage{bbold}
\usepackage{hyperref}

\usepackage{relsize}
\def\Babar{{\mbox{\slshape B\kern-0.1em{\smaller A}\kern-0.1em B\kern-0.1em{\smaller A\kern-0.2em R}}}}

\newcommand{\lsim}{
	\mathrel{\hbox{\rlap{\hbox{\lower4pt\hbox{$\sim$}}}\hbox{$<$}}}}

\newcommand{\gsim}{
	\mathrel{\hbox{\rlap{\hbox{\lower4pt\hbox{$\sim$}}}\hbox{$>$}}}}

\renewcommand{\arraystretch}{2}

\allowdisplaybreaks[2]

\newcommand{\nn}{\nonumber}
\newcommand{\ord}{{\cal O}}

\def\mLb{{m_{\Lambda_b}}}
\def\mmLb{{m^2_{\Lambda_b}}}
\def\mL{{m_{\Lambda}}}
\def\mmL{{m^2_{\Lambda}}}
\def\cl{{\cos\theta_\ell}}

\def\sl{{\sin\theta_\ell}}

\def\plpl{{+\frac{1}{2}+\frac{1}{2}}}
\def\plmi{{+\frac{1}{2}-\frac{1}{2}}}
\def\mipl{{-\frac{1}{2}+\frac{1}{2}}}
\def\mimi{{-\frac{1}{2}-\frac{1}{2}}}
\def\re{{\rm Re}}  
\def\mC{{\mathcal{C}}}

\def\bl{{\beta_\ell}}

\def\blp{{\beta^\prime_\ell}}

\definecolor{schrift}{RGB}{120,0,0}

\def\mLQ{m_{\rm LQ}^2}
\def\mLQQ{m_{\rm LQ}^4}

\def\ARpe#1{{A^R_{\perp_{#1}}}}  \def\ARpa#1{{A^R_{\|_{#1}}}}

\def\ALpe#1{{A^L_{\perp_{#1}}}}  \def\ALpa#1{{A^L_{\|_{#1}}}}

\def\APpa{{A_{\rm P \|}}} \def\APpe{{A_{\rm P \perp}}}
\def\AsPpa{{A^\ast_{\rm P \|}}} \def\AsPpe{{A^\ast_{\rm P \perp}}}
\def\ASpa{{A_{\rm S \|}}} \def\ASpe{{A_{\rm S \perp}}}
\def\AsSpa{{A^\ast_{\rm S \|}}} \def\AsSpe{{A^\ast_{\rm S \perp}}}

\def\Apat{{A_{\|t}}} \def\Apet{{A_{\perp t}}}

\def\AsRpe#1{{A^{R \ast}_{\perp_{#1}}}}  \def\AsRpa#1{{A^{R \ast}_{\|_{#1}}}}

\def\AsLpe#1{{A^{L \ast}_{\perp_{#1}}}}  \def\AsLpa#1{{A^{L \ast}_{\|_{#1}}}}

\def\bd0tau{B\to D \tau\nu_{\tau}}

\def\be {\begin{equation}}
\def\ee {\end{equation}}

\usepackage[normalem]{ulem}

\definecolor{darkgreen}{cmyk}{1,0,1,0.4}

\definecolor{pink}{cmyk}{0.4,1,0.3,0}

\def\com2#1{\textcolor{red}{\it{#1}}}
\begin{document}
	
	%opening
\title{Lepton flavor violating $\Lambda_b\to\Lambda\ell_1\ell_2$ decay}
	
\author{Diganta Das}
\email{diganta99@gmail.com}
\affiliation{ Department of Physics and Astrophysics, University of Delhi, Delhi 110007, India}

\begin{abstract}  
Inspired by the recent hints of lepton flavor universality violation in $b\to s\ell\ell$ and $b\to c\ell\nu$ transitions, we study lepton flavor violating exclusive $\Lambda_b\to\Lambda\ell_1^+\ell_2^-$ ($\ell_1\neq \ell_2$) decay, which is forbidden in the Standard Model. Starting from a general effective Hamiltonian for a $b\to s\ell_1^+\ell_2^-$ transition that includes vector and axial-vector operators, and scalar and pseudo-scalar operators, we derive a two-fold decay distribution of $\Lambda_b\to\Lambda\ell_1^+\ell_2^-$. The distribution helps us to construct the differential branching ratio and the lepton side forward-backward asymmetry, which are studied in a vector leptoquark model. The parameter space of the vector leptoquark model is constrained by low energy observables. 
\end{abstract}   
	
\maketitle

%%%%%%%%%%%%%%%%%%%%%%%%%%%%%%%%%%%%%%%%%%%%%%%%%%%%%%%%%%%%%%%%%%%%%%
%%%%%%%%%%%%%%%%%%%%%%%%%%%%%%%%%%%%%%%%%%%%%%%%%%%%%%%%%%%%%%%%%%%%%%
\section{Introduction}
Though an $\ord(1)$ signal of new physics (NP) is still at large, the recent results by the Belle and LHCb Collaborations in the neutral and charged current transitions of $b$-flavored mesons are intriguing hints of lepton flvor universality (LFU) violation, which is absent in the Standard Model (SM). In the flavor changing neutral current transition $b\to s\ell\ell$ the observables that probe LFU are
\begin{equation}
R_{K^{(\ast)}} = \frac{\mathcal{B}(B\to K^{(\ast)}\mu^+\mu^-)}{\mathcal{B}(B\to K^{(\ast)} e^+ e^-)}\, .
\end{equation}
The LHCb Collaboration has measured $R_K$ and the most recent result is \cite{Aaij:2019wad}
\begin{equation}
	R_K = 0.846^{+0.060+0.016}_{-0.054-0.014}\, , 1 \le q^2 \le 6.0 {\rm GeV}^2\, ,
\end{equation}
where $q^2$ is the invariant mass squared of the final state dilepton pair. This result is lower than the SM prediction $R_K^{\rm SM}=1.00\pm 0.01$ \cite{Bordone:2016gaq} by about $2.5\sigma$. On the other hand, the most recent measurements of $R_{K^{(\ast)}}$ by the LHCb \cite{Aaij:2017vbb} in the two dilepton invariant mass squared bins
\begin{equation}
R_{K^\ast} = \begin{cases}
0.66^{0.11}_{-0.17} \pm 0.03, \!\!\!\quad 0.045 \le q^2 \le 1.1 {\rm GeV}^2\, , \\
0.69^{0.11}_{-0.07} \pm 0.05, \!\!\!\quad 1.1 \le q^2 \le 6.0 {\rm GeV}^2\, ,
\end{cases} 
\end{equation}
deviate from the SM predictions $R_{K^\ast}^{\rm SM} =0.906 \pm 0.028$ and $R_{K^\ast}^{\rm SM} = 1.00 \pm 0.01$ by 2.3$\sigma$ and $2.5\sigma$, respectively. Belle has also presented \cite{Abdesselam:2019wac} their results of $R_K$ and $R_{K^\ast}$ which are closer to the SM predictions but has large uncertainties.

Independently of the results in the $b\to s\ell\ell$ transitions, hints of LFU violation have also been found in the charged current transition $b\to c\ell\nu$. The observables in which deviations from the SM predictions have been observed are $R_D$ and $R_{D^\ast}$,
\begin{equation}
R_{D^{(\ast)}} = \frac{\mathcal{B}(B\to D^{(\ast)}\tau\nu)}{\mathcal{B}(B\to D^{(\ast)}\ell\nu)}\, ,\quad \ell = e, \mu.
\end{equation}	
$R_{D^{\ast}}$ has been measured by Belle \cite{Huschle:2015rga,Abdesselam:2016cgx,Hirose:2016wfn} LHCb \cite{Aaij:2015yra} and BaBar \cite{Lees:2013uzd}. The new measurement by Belle \cite{Abdesselam:2019dgh} using semi-leptonic tagging gives
\begin{align}
& R_D = 0.307 \pm 0.37 \pm 0.016\, ,\\
& R_{D^\ast} = 0.283 \pm 0.018 \pm 0.14\, .
\end{align}
HFLAV has combined the most recent results and their averages \cite{Amhis:2016xyh} exceed the SM predictions $R_D^{\rm SM} = 0.299 \pm 0.003$ \cite{Bigi:2016mdz} and $R_{D^\ast}^{\rm SM} = 0.258 \pm 0.005$ \cite{Jaiswal:2017rve} by $2.3\sigma$ and $3.4\sigma$, respectively. 

A number of NP models with new particle content has been constructed that can explain these deviations. Shortly after the first hints of LFU violation were announced \cite{Aaij:2014ora} it was shown in Ref.~\cite{Glashow:2014iga} that LFU violation implies lepton flavor violating (LFV) interactions. Despite several counter examples to this observation \cite{Celis:2015ara, Alonso:2015sja}, most models that generate LFU violation also generate LFV processes which are strictly forbidden in the SM. Therefore, the observation of LFV decay will be a smoking gun signal of NP. Some of the LFV processes that have been extensively looked for are leptonic decays $\tau\to 3\mu$, $\mu\to 3e$ etc and $\ell \to \ell^\prime M$ where $M$, is a meson, radiative decays $\mu\to e\gamma$ etc, and $\mu\to e$ conversion. Interestingly, in the Higgs sector $h\to \mu\tau$ was studied and an apparent excess was also reported by CMS \cite{Khachatryan:2015kon}, which disappeared in subsequent measurements. 

In this paper we discuss LFV baryonic decay $\Lambda_b\to\Lambda\ell_1^+\ell_2^-$, which proceeds through a $b\to s\ell_1^+\ell_2^-$ transition where $\ell_1^+$ and $\ell_2^-$ are charged leptons of different flavors. Though its SM counterpart $\Lambda_b\to\Lambda\ell\ell$ has been measured by the LHCb \cite{Aaij:2015xza}\cite{Aaij:2018gwm}, to the best of our knowledge currently there are no experimental data on $\Lambda_b\to\Lambda\ell_1^+\ell_2^-$. Unlike $\Lambda_b\to\Lambda\ell\ell$, the advantage with $\Lambda_b\to\Lambda\ell_1^+\ell_2^-$ decay is that it does not suffer from long-distance QCD and charmonium resonance backgrounds. The $\Lambda_b\to\Lambda\ell_1^+\ell_2^-$ decay was earlier discussed in \cite{Sahoo:2016nvx} in the context of scalar leptoquark model where only vector and axial-vector type effective operators were considered. In this paper we include in addition scalar and pseudo-scalar operators and present a double differential distribution. From this distribution we study the differential branching ratio and the forward-backward asymmetry. These observables are studied in a vector leptoquark model $U_1\equiv(\textbf{3,1})_{2/3}$. We use several low energy observables to constrain the model parameters.

The paper is organized as follows. We begin by describing in Sect.~\ref{subsec:effHam} the effective Hamiltonian for a $b\to s\ell_1^+\ell_2^-$ transition. The differential decay distribution of the exclusive $\Lambda_b\to\Lambda\ell_1^+\ell_2^-$ is calculated in Sect.~\ref{sec:angular} followed by a numerical analysis in Sect.~\ref{sec:num}. We summarize our discussions in Sect.~\ref{sec:summary}.

%%%%%%%%%%%%%%%%%%%%%%%%%%%%%%%%%%%%%%%%%%%%%%%%%%%%%%%%%%%%%%%%%%%%%%%%%%%%%%%%%%%%%%%%%%%%%%%%%%%%%%%%%%%%%%%%%%%%%%%%%%%%%%%%%%%%%%%%%%%%%%%%
%\section{Exclusive $\Lambda_b \to\Lambda\ell_1\ell_2$ decay }
\section{Effective Hamiltonian \label{subsec:effHam}}
We start with the following effective Hamiltonian for the lepton flavor violating $b\to s\ell_1^+\ell_2^-$ transition:
\begin{equation}\label{eq:Heff1}
\mathcal{H}^{\rm eff} = - \frac{1}{2v^2}V_{tb}V_{ts}^\ast\frac{\alpha_e}{4\pi}  \sum_i\bigg( \mC_i \mathcal{O}_i +  \mC^\prime_i \mathcal{O}^\prime_i \bigg)\, ,
\end{equation}
where $v^2=1/(\sqrt{2}G_F)\approx 246$ GeV is the SM vacuum expectation value, and $i = V, A, S, P$ correspond to vector, axial-vector, scalar, and pseudo-scalar operators, which read
\begin{eqnarray}\label{eq:opbasis}
\begin{split}
&\mathcal{O}^{(\prime)}_V = \big[\bar{s}\gamma^\mu P_{L(R)}b \big]\big[\ell_2\gamma_\mu\ell_1 \big]\, ,\\
&\mathcal{O}^{(\prime)}_{A} = \big[\bar{s}\gamma^\mu P_{L(R)}b \big]\big[\ell_2\gamma_\mu\gamma_5\ell_1 \big]\, ,\\
&\mathcal{O}_S^{(\prime)} = \big[\bar{s}P_{R(L)}b \big]\big[\ell_2\ell_1 \big]\, ,\quad \mathcal{O}_P^{(\prime)} = \big[\bar{s}P_{R(L)}b \big]\big[\ell_2\gamma_5\ell_1 \big]\, .
\end{split}
\end{eqnarray}
Here $\alpha_e$ is the fine structure constant, $V_{tb}V_{ts}^\ast$ are the Cabibbo--Kobayashi--Maskawa matrix elements, $P_{L,R}=(1\mp\gamma_5)/2$ are the chiral projectors. The $\mC_{V,A,S,P}^{(\prime)}$ are the short-distance Wilson coefficients that vanish in the SM but can be non-zero in many scenarios beyond the SM. In the SM $\ell_1,\ell_2$ are leptons of the same flavor, say $\ell$, and it is customary to denote the operators $\mathcal{O}_{V,A}$ as $\mathcal{O}_{9,10}$ with the corresponding Wilson coefficients $\mC_{9,10}$. Additionally, in the SM there is also a dipole operator $\mathcal{O}_7$ that contributes to the $b\to s\ell\ell$ transition. The long-distance part of the decay is encoded in the $\Lambda_b\to\Lambda$ transition matrix elements (see \cite{Das:2018iap} for definitions) which are parametrized in terms of six $q^2$ dependent form factors $f^V_{t,0,\perp}$, $f^A_{t,0,\perp}$ \cite{Feldmann:2011xf}. For our numerical analysis the form factors are taken from calculations in lattice QCD \cite{Detmold:2016pkz}.

\section{Exclusive $\Lambda_b \to\Lambda\ell_1^+\ell_2^-$ decay \label{sec:angular}}
To set up the kinematics of the decay we assume that the $\Lambda_b$ is at rest while the $\Lambda$ and the dilepton pair travel along the $+z$- and $-z$-axis, respectively. We assign $p,k,q_1$ and $q_2$ as the momenta of the $\Lambda_b, \Lambda, \ell_1$, and $\ell_2$, and $s_p, s_k$ are the spins of $\Lambda_b, \Lambda$ on to the $z-$axis in their respective rest frames. We also introduce two kinematic variables; $q^\mu = q_1^\mu + q_2^\mu$  is the four-momentum of the dielpton pair, and $\theta_\ell$ is the angle that the lepton $\ell_1$ makes with respect to the $z$-axis in the dilepton rest frame. The decay amplitudes can be written as
\begin{align}\label{eq:Mll}
& \mathcal{M}^{\lambda_2,\lambda_1}(s_p,s_k) = - \frac{V_{tb}V_{ts}^\ast}{2v^2}\frac{\alpha_e}{4\pi} \sum_{i=L,R}\bigg[ \sum_{\lambda} \eta_\lambda H^{i,s_p,s_k}_{\rm VA, \lambda} L^{\lambda_2,\lambda_1}_{i,\lambda}\, \nn\\& + H^{i,s_p,s_k}_{\rm SP} L^{\lambda_2,\lambda_1}_i  \bigg]\, .
\end{align}
Here $H^{i,s_p,s_k}_{\rm VA, \lambda}$ and $H^{i,s_p,s_k}_{\rm SP}$ are the hadronic helicity amplitudes corresponding to vector and axial-vector (VA), and scalar and pseudo-scalar (SP) operators, and the $L^{\lambda_2,\lambda_1}_{i,\lambda}, L^{\lambda_2,\lambda_1}_i$ are the leptonic helicity amplitudes. Here $i={ L,R}$ corresponds to the chiralities of the lepton current and the $\lambda=t,\pm 1,0$ are the helicity states of the virtual gauge boson that decay into the dilepton pair. The $\lambda_{1,2}$ are the helicities of the leptons and $\eta_t=1,\eta_{\pm 1,0}=-1$. The definitions and the expressions of of $H^{i,s_p,s_k}_{\rm VA, \lambda}$ and $H^{i,s_p,s_k}_{\rm SP}$ in terms of Wilson coefficients and form factors can be found in \cite{Das:2018sms}.  In the literature, instead of the hadronic helicity amplitudes, transversity amplitudes $A^i_{\perp(\|)_{\rm 1}}, A^i_{\perp(\|)_{\rm 0}}$ and $A_{\rm S\perp(\|)}, A_{\rm P\perp(\|)}$ are often used. Following \cite{Das:2018iap} the expressions of the transversity amplitudes are collected in Appendix~\ref{sec:TAs2}. 

The $L^{\lambda_2,\lambda_1}_{i,\lambda}$ and $L^{\lambda_2,\lambda_1}_{i}$ amplitudes are defined as
\begin{align}
\begin{split}
& L^{\lambda_2,\lambda_1}_{L(R)} = \langle \bar{\ell}_2(\lambda_2)\ell_1(\lambda_1) | \bar{\ell}_2 (1\mp\gamma_5) \ell_1 | 0\rangle\, , \\
\label{eq:Ldef2}
& L^{\lambda_2,\lambda_1}_{L(R),\lambda} = \bar{\epsilon}^\mu(\lambda) \langle \bar{\ell}_2(\lambda_2) \ell_1(\lambda_1) | \bar{\ell}_2 \gamma_\mu (1\mp\gamma_5) \ell_1 | 0\rangle\, ,
\end{split}
\end{align}
where $\epsilon^{\mu}$ is the polarization vector of the virtual gauge boson that decays in to the dilepton pair. The details of the calculations of $L^{\lambda_2,\lambda_1}_{i,\lambda}$ and $L^{\lambda_2,\lambda_1}_{i}$ are given in Appendix~\ref{sec:llRF}. Based on these calculations we obtain the differential branching ratio of $\Lambda_b\to\Lambda\ell_1\ell_2$ as 
\begin{equation}\label{eq:twofold}
\frac{d\mathcal{B}}{dq^2 d\cos\theta_\ell} = \frac{3}{2} \bigg( K_{1ss} \sin^2\theta_\ell + K_{1cc} \cos^2\theta_\ell + K_{1c} \cos\theta_\ell \bigg)\, .
\end{equation}
Each of the angular coefficients $K_{1ss,1cc,1c}$ can be written in the following way:
\begin{equation}
K_{1ss, 1cc} = K_{1ss, 1cc}^{\rm VA} + K_{1ss, 1cc}^{\rm SP} +K_{1ss, 1cc}^{\rm int}\, ,
\end{equation} 
where $K_{1ss, 1cc,1c}^{\rm VA}, K_{1ss, 1cc,1c}^{\rm SP}$ are contributions from VA and SP operators, and $K_{1ss, 1cc,1c}^{\rm int}$ includes their interference terms. In terms of the transversity amplitudes the expressions of $K_{1ss, 1cc,1c}^{\rm VA}, K_{1ss, 1cc,1c}^{\rm SP}$ read
\begin{align}
&K_{1ss}^{\rm VA} = \frac{1}{4} \bigg( 2|\ARpa0|^2 + |\ARpa1|^2 + 2|\ARpe0|^2 + |\ARpe1|^2 +  \nn\\& \{ R \leftrightarrow L  \} \bigg)- \frac{m_+^2+m_-^2}{4q^2} \bigg[ \bigg( |A^R_{\|_0}|^2 + |A^R_{\perp_0}|^2 + \{ R \leftrightarrow L \} \bigg) \, \nn\\ &- \bigg( |A_{\perp t}|^2  + \{ \perp \leftrightarrow \| \}\bigg) \bigg]  + \frac{m_+^2-m_-^2}{4q^2} \bigg[ 2\re\bigg( A^R_{\perp_0}A^{\ast L}_{\perp_0}\, \nn\\& + A^R_{\perp_1}A^{\ast L}_{\perp_1} + \{ \perp \leftrightarrow \| \}  \bigg) \bigg] - \frac{m_+^2m_-^2}{4q^4} \bigg[ \bigg(|\ARpa1|^2 + |\ARpe1|^2 \nn\\&+ \{ R \leftrightarrow L \} \bigg) + 2|A_{\|t}|^2 + 2|A_{\perp t}|^2\bigg] \, , \\
%
%\end{align}
%\begin{align}
%
&K_{1cc}^{\rm VA} = \frac{1}{2}\bigg( |\ARpa1|^2 + |\ARpe1|^2 + \{R \leftrightarrow L \} \bigg) + \frac{m_+^2+m_-^2}{4q^2}\,\nn\\&\times \bigg[ \bigg( |A^R_{\|_0}|^2 - |A^R_{\|_1}|^2 + |A^R_{\perp_0}|^2 - |A^R_{\perp_1}|^2 + \{ R \leftrightarrow L \} \bigg)\,\nn\\ &+ \bigg( |A_{\perp t}|^2 + |A_{\| t}|^2 \bigg) \bigg] + \frac{m_+^2-m_-^2}{4q^2} \bigg[ 2\re\bigg( A^R_{\perp_0}A^{\ast L}_{\perp_0}\, \nn\\& + A^R_{\perp_1}A^{\ast L}_{\perp_1} + \{\perp \leftrightarrow \| \} \bigg) \bigg] - \frac{m_+^2 m_-^2}{2q^4}  \bigg[ \bigg(|\ARpa0|^2 + |\ARpe0|^2 \nn\\&+ \{ R \leftrightarrow L \} \bigg) + |A_{\|t}|^2 + |A_{\perp t}|^2\bigg] \, , \\
%
%\end{align}
%\begin{align}
%
&K_{1c}^{\rm VA} = -\beta_\ell \beta_\ell^\prime \bigg( A^R_{\perp_1}A^{\ast R}_{\|_1} - \{ R \leftrightarrow L \}  \bigg)\,\nn\\& + \beta_\ell \beta_\ell^\prime\frac{m_+ m_-}{q^2}  \re\bigg( \ALpa0 A_{\| t}^\ast + \ALpe0 A_{\perp t}^\ast \bigg)  \, ,\\
%\end{align}
%
%\begin{align}
&K_{1ss}^{\rm SP} = \frac{1}{4}\bigg( |A_{\rm S\perp}|^2 + |A_{\rm P\perp}|^2 + \{ \perp \leftrightarrow \| \} \bigg) \,\nn\\&- \frac{m_+^2}{4q^2}\big(|A_{S \|}|^2 + |A_{S \perp}|^2\big) - \frac{m_-^2}{4q^2}\big(|A_{P \|}|^2 + |A_{P \perp}|^2\big) \, ,~~~~\\
&K_{1cc}^{\rm SP} = \frac{1}{4}\bigg( |A_{\rm P\perp}|^2 + |A_{\rm S\perp}|^2 + \{\perp \leftrightarrow \| \} \bigg)\, \nn\\& - \frac{m_+^2}{4q^2}\big(|A_{S \|}|^2 + |A_{S \perp}|^2\big) - \frac{m_-^2}{4q^2}\big(|A_{P \|}|^2 + |A_{P \perp}|^2\big)\, ,\\
&K_{1c}^{\rm SP} = 0\, .
\end{align}

The interference terms read
\begin{align}
&K_{1ss}^{\rm int} = \frac{m_+}{2\sqrt{q^2}} \re\bigg(\Apat\AsPpa + \Apet\AsPpe  \bigg) + \,\nn\\&\frac{m_-}{2\sqrt{q^2}} \re\bigg( \Apat\AsSpa + \Apet\AsSpe \bigg)- \frac{m_+^2m_-}{2q^2\sqrt{q^2}} \re\bigg( \Apat\AsSpa + \, \nn\\& \Apet\AsSpe \bigg) - \frac{m_+m_-^2}{2q^2\sqrt{q^2}}\re\bigg( \Apat\AsPpa +  \Apet\AsPpe \bigg)\, ,\\
&K_{1cc}^{\rm int} = \frac{m_+}{2\sqrt{q^2}}\re\bigg(\Apat\AsPpa + \Apet\AsPpe \bigg) + \frac{m_-}{2\sqrt{q^2}}\re\bigg(\Apat\AsSpa \,\nn\\&+ \Apet\AsSpe \bigg) - \frac{m_+^2m_-}{2q^2\sqrt{q^2}}\re\bigg( \Apat\AsSpa + \Apet\AsSpe \bigg)\,\nn\\& - \frac{m_+ m_-^2}{2q^2\sqrt{q^2}}\re\bigg( \Apat\AsPpa + \Apet\AsPpe \bigg)\, ,\\
&K_{1c}^{\rm int} = \frac{\beta_\ell\beta_\ell^\prime}{2\sqrt{q^2}} \re\bigg( \ASpa\AsLpa{0} + \ASpe\AsLpe{0} + \ASpa\AsRpa{0} + \ASpe\AsRpe{0} \bigg) \nn\\ &+ \frac{\beta_\ell\beta_\ell^\prime}{2\sqrt{q^2}} \re\bigg( \APpa\AsLpa{0} + \APpe\AsLpe{0} - \APpa\AsRpa{0} - \APpe\AsRpe{0} \bigg) 
\end{align} 
We have defined $m_\pm = m_1 \pm m_2$ where $m_1, m_2$ are the masses of $\ell_1, \ell_2$, respectively, and the factors $\beta_{\ell}^{(\prime)}$ are defined in Appendix \ref{sec:TAs2}. From the differential decay distribution we define two observables \cite{Das:2018iap}; differential branching ratio 
\begin{equation} \label{eq:diffBr}
	\frac{d\mathcal{B}}{dq^2} = 2 K_{1ss} + K_{1cc}\, ,
\end{equation}
and the forward backward asymmetry 
\begin{equation}\label{eq:AlFB}
A^{\ell}_{\rm FB} = \frac{3}{2} \frac{K_{1c}}{K_{1ss} + K_{1cc}}\, .
\end{equation}
The available phase space in the dilepton invariant mass squared $q^2$ is
\begin{equation}
	(m_1+m_2)^2 \le q^2 \le (\mLb - \mL)^2\, .
\end{equation}

%%%%%%%%%%%%%%%%%%%%%%%%%%%%%%%%%%%%%%%%%%%%%%%%%%%%%%%%%%%%%%%%%%%%%
%%%%%%%%%%%%%%%%%%%%%%%%%%%%%%%%%%%%%%%%%%%%%%%%%%%%%%%%%%%%%%%%%%%%%
\section{Numerical analysis \label{sec:num}}
%%%%%%%%%%%%%%%%%%%%%%%%%%%%%%%%%%%%%%%%%%%%%%%%%%%%%%%%%%%%%%%%%%%%%%
Among many leptoquark models proposed to explain flavor anomalies, the vector leptoquark $U_1\equiv({\bf 3},{\bf 1})_{2/3}$ has emerged as an excellent candidate that can simultaneously alleviate the tensions between theory and experiments in both the charged and the neutral current sectors. In fact, $U_1$ can accommodate a large number of low energy data and high-$p_T$ searches without too much fine-tuning of the model parameters \cite{Buttazzo:2017ixm}. Early works reconciling these anomalies by coupling the $U_1$ with the third generation quarks and leptons can be found in Refs. \cite{Alonso:2015sja, Barbieri:2015yvd}. The UV completion of this model has also recently been discussed in Ref.~\cite{Cornella:2019hct}. The SM gauge symmetry allows couplings of the $U_1$ leptoquark to both left- and right-handed fermions and the Lagrangian reads
\begin{equation}\label{eq:U1}
\mathcal{L} \supset \frac{U^\mu_1}{\sqrt{2}} \bigg[ \beta^{ij}_L (\bar{Q}^i_L \gamma_\mu L^j_L ) + \beta^{ij}_R (\bar{d}^i_R \gamma_\mu \ell^j_R ) \bigg]\, .
\end{equation}
Here the $Q^i_L = (V^\ast_{ji}u^j_L~d^i_L)^T$ and $L^i_L = (\nu^i_L~\ell^i_L)^T$ are $SU(2)_L$ doublets, and the $\beta_{L,R}$ are $3\times 3$ Yukawa matrices. To address the flavor anomalies we assume the following flavor ansatz:
\begin{equation}\label{eq:R2flav}
\beta_L = 
\begin{pmatrix}
0 && 0 && 0 \\
0 && \beta_L^{s\mu} && \beta_L^{s\tau} \\
0 && \beta_L^{b\mu} && \beta_L^{b\tau}
\end{pmatrix}\, ,\quad
\beta_R = 
\begin{pmatrix}
0 && 0 && 0 \\
0 && 0 && 0 \\
0 && 0 && \beta_R^{b\tau}
\end{pmatrix}\, .
\end{equation}
With the couplings to the first generation set to zero the experimental limits on atomic parity violation, $\mu-e$ conversion on nuclei, and $\mathcal{B}(K\to \pi\bar{\nu}\nu)$ are evaded. An important feature of the vector leptoquark model is the absence of the tree level $b\to s\nu\bar{\nu}$ transition evading the current experimental constraints coming from $B\to K^\ast\nu\bar{\nu}$ \cite{Grygier:2017tzo}. There is also a ``flavor protection'' mechanism in the $U_1$ loops due to which the purely leptonic processes $\tau\to 3\mu$, $\tau\to \mu\nu\bar{\nu}$ and $b\to s\nu\bar{\nu}$ have little phenomenological significance \cite{DiLuzio:2018zxy, Buttazzo:2017ixm, Crivellin:2018yvo}. These processes aside, we consider a number of low energy flavor observables to constrain the flavor structure \eqref{eq:R2flav}. 

The Lagrangian \eqref{eq:U1} generates the following VA and SP operators for $b\to s\ell_1^+\ell_2^-$:
\begin{align}\label{eq:VLQWC}
& \mC^{\ell_1\ell_2}_V = - \mC_A^{\ell_1\ell_2} = - \frac{\pi v^2}{2V_{tb}V_{ts}^\ast \alpha_e \mLQ} \beta^{s\ell_2}_L (\beta^{b\ell_1}_L)^\ast\,, \\
& \mC^{\prime\ell_1\ell_2}_V = \mC_A^{\prime\ell_1\ell_2} = -\frac{\pi v^2}{2V_{tb}V_{ts}^\ast \alpha_e \mLQ} \beta^{s\ell_2}_R (\beta^{b\ell_1}_R)^\ast\,,\\
& \mC^{\ell_1\ell_2}_S = - \mC_P^{\ell_1\ell_2} = \frac{\pi v^2}{V_{tb}V_{ts}^\ast \alpha_e \mLQ} \beta^{s\ell_2}_L (\beta^{b\ell_1}_R)^\ast\,,\\
& \mC^{\prime\ell_1\ell_2}_S = \mC_P^{\prime\ell_1\ell_2} = \frac{\pi v^2}{V_{tb}V_{ts}^\ast \alpha_e \mLQ} \beta^{s\ell_2}_R (\beta^{b\ell_1}_L)^\ast\,.
\end{align}
For the given flavor ansatz \eqref{eq:R2flav} $R_{K^{(\ast)}}$ receives the following modifications \cite{Celis:2017doq} through the NP Wilson coefficients $\mC_{V,A}^{\mu\mu}$:
\begin{align}
& R_K^{[1,6]{\rm GeV}^2} \approx 1 + 0.46 C_V^{\mu\mu}\, ,\\
& R_{K^\ast}^{[1.1,6]{\rm GeV}^2} \approx 1 + 0.47 C_V^{\mu\mu}\, .
\end{align}
Global fits to the most recent $b\to s\mu\mu$ data have been performed by several groups and we take the range $-0.59\le\mC_V^{\mu\mu}=-\mC_A^{\mu\mu}\le-0.40$ \cite{Alguero:2019ptt} \cite{Aebischer:2019mlg} in our analysis. For a large $\beta_L^{s\tau}$ there are additional flavor-universal contributions to the $\to s\ell\ell$ in the direction of $\mC_V^{\mu\mu}$ due to the off-shell photon penguins \cite{Crivellin:2018yvo}
\begin{equation}
	\Delta \mC_V \approx  -\frac{v^2}{6\mLQ V_{tb}V_{ts}^\ast} \beta_L^{s\tau} (\beta_L^{s\tau})^\ast \log\bigg(\frac{m_b^2}{\mLQ}\bigg)\, .
\end{equation}
Experiments yield $\Delta \mC_V =-0.73 \pm 0.23$ \cite{Alguero:2019ptt} \cite{Aebischer:2019mlg}.

While the contributions of $U_1$ leptoquark to $b\to s\mu\mu$ processes are through vector and axial-vector operators only, in the presence of a right-handed coupling $\beta_R^{b\tau}$ scalar and pseudo-scalar currents can contribute to $b\to s\tau\tau$ processes $B_s\to\tau^+\tau^-$ and $B\to K\tau^+\tau^-$. The $B_s\to\tau\tau$ branching ratio reads 
\begin{align}
&\mathcal{B}(B_s\to\tau^+\tau^-) = \mathcal{B}(B_s\to\tau^+\tau^-)_{\rm SM}\, \nn\\&\times\bigg|1 + \frac{\pi v^2}{2 V_{tb}V_{ts}^\ast \alpha \mLQ} \frac{\beta_L^{s\tau} }{C_{10}^{\rm SM}} \bigg(\beta_L^{b\tau} - \frac{ m_{B_s}^2}{m_\tau(m_s+m_b)} (\beta_R^{b\tau})^\ast \bigg)  \bigg|^2\, \nn\\& + \bigg(1-\frac{4m_\tau^2}{m^2_{B_s}} \bigg) \bigg| \frac{\pi v^2}{2 V_{tb}V_{ts}^\ast \alpha \mLQ C_{10}^{\rm SM}} \frac{ m_{B_s}^2 \beta_L^{s\tau}(\beta_R^{b\tau})^\ast}{m_\tau(m_s+m_b)}  \bigg|^2\,.
\end{align}
The present experimental upper limit is $\mathcal{B}(B_s\to\tau^+\tau^-)<0.0(3.4)\times 10^{-3}$ \cite{Aaij:2017xqt} and the SM prediction read $\mathcal{B}(B_s\to\tau^+\tau^-)<(7.73\pm 0.49)\times 10^{-7}$ \cite{Bobeth:2013uxa}. The SM branching ratio of $B\to K\tau^+\tau^-$ is $\mathcal{B}(B\to K\tau^+\tau^-)=1.44(0.28)\times 10^{-7}$ where we use hadronic inputs from \cite{Bouchard:2013pna}, and the experimental upper bound is $\mathcal{B}(B\to K\tau^+\tau^-)=(1.36\pm 0.71)\times 10^{-3}$ \cite{TheBaBar:2016xwe}. 

The leptoquark also contributes to the LFV observables $\mathcal{B}(B^+\to K^+\tau^{\pm}\mu^{\mp})$ and $\mathcal{B}(\tau\to\mu\phi)$. Following the simplified expressions given in \cite{Bordone:2018nbg} we get 
\begin{align}
& \mathcal{B}(B^+\to K^+\tau^+\mu^-) \approx \frac{v^4}{\mLQQ}\bigg( 0.50\big|\beta_L^{s\mu}(\beta_L^{b\tau})^\ast\big|^2\, \nn\\& + 2.83 \big|\beta_L^{s\mu}(\beta_R^{b\tau})^\ast\big|^2 - 1.39 \re[\beta_L^{b\tau} (\beta_R^{b\tau})^\ast]|\beta_L^{s\mu}|^2 \bigg)\, ,\\
& \mathcal{B}(B^+\to K^+\tau^-\mu^+) \approx \frac{v^4}{\mLQQ}0.50 \big|\beta_L^{b\mu}(\beta_L^{s\tau})^\ast\big|^2 \, .
\end{align}
The experimental upper limit is $\mathcal{B}(B^+\to K^+\tau^+\mu^-) \leq 2.8\times 10^{-5}$ and $\mathcal{B}(B^+\to K^+\mu^+\tau^-) \leq 4.5\times 10^{-5}$ \cite{Lees:2012zz}. For the $\tau\to \mu\phi$ decay, following \cite{Goto:2010sn} we get after neglecting the mass of the muon 
\begin{align}
\mathcal{B}(\tau\to\mu\phi) &= \frac{f_\phi^2 m_\tau^3}{32\pi\Gamma_\tau }\cdot \frac{1}{16\mLQQ} \bigg(1-\frac{m^2_\phi}{m^2_\tau} \bigg)^2 \bigg(1+2\frac{m^2_\phi}{m^2_\tau} \bigg)\, \nn\\&\times \big|\beta_L^{s\tau} (\beta_L^{s\mu})^\ast \big|^2\, .
\end{align}
The experimental upper limit from Belle \cite{Miyazaki:2011xe} is $\mathcal{B}(\tau\to\mu\phi)\leq(0.0\pm 5.1)\times 10^{-8}$. In the presence of right-handed coupling, $\tau\to\mu\gamma$ is also induced: 
\begin{equation}
\mathcal{B}(\tau\to\mu\gamma) = \frac{1}{\Gamma_\tau}\frac{\alpha_e}{64\pi^4} \frac{m_\tau^3m_b^2}{16\mLQQ} |\beta_R^{b\tau}(\beta_L^{b\mu})|^2\, .
\end{equation}
The experimental upper bound is $\mathcal{B}(\tau\to\mu\gamma)=0.0(3.0)\times 10^{-8}$ \cite{Amhis:2016xyh}.

The charged current transition $b\to c\ell\nu$ also receives contributions from the vector leptoquark. Here the flavor of the final state neutrino in general may be different from the flavor of the accompanying lepton. The most general effective Hamiltonian for this transition is
\begin{align}
\mathcal{H}^{b\to c\ell\bar{\nu}}_{\rm eff} &= \frac{2 V_{cb}}{v^2} \bigg( \big(1+\mC_{V_1}^{\ell}\big) \mathcal{O}_{V_1} + \mC_{V_2}^{\ell}\mathcal{O}_{V_2}\,\nn\\& + \mC_{S_1}^{\ell}\mathcal{O}_{S_1} + \mC_{S_2}^{\ell}\mathcal{O}_{S_2} + \mC_T^{\ell}\mathcal{O}_T   \bigg)\, ,
\end{align}
where the operators are given by
\begin{align}
&\mathcal{O}_{V_1} = (\bar{c}_L\gamma^\mu b_L)(\bar{\ell}_L\gamma_\mu \nu_{L})\, ,\quad \mathcal{O}_{V_2} = (\bar{c}_R\gamma^\mu b_R)(\bar{\ell}_L\gamma_\mu \nu_{L})\, ,\nn\\
&\mathcal{O}_{S_1} = (\bar{c}_L b_R)(\bar{\ell}_R \nu_{L})\, ,\quad \mathcal{O}_{S_2} = (\bar{c}_R b_L)(\bar{\ell}_R \nu_{L})\, ,\\
&\mathcal{O}_T = (\bar{c}_R\sigma^{\mu\nu}b_L)(\bar{\ell}_R\sigma_{\mu\nu}\nu_{L})\, .\nn
\end{align}
In the SM all the Wilson coefficients $\mC^{\ell}_{i} =0$. In the $U_1$ leptoquark model the only non-vanishing Wilson coefficients are 
\begin{align}
&\mC^{\ell}_{V_1} = \frac{v^2}{4\mLQ} (\beta^{b\ell}_L)^\ast \bigg(\beta_L^{b\ell} + \frac{V_{cs}}{V_{cb}} \beta_L^{s\ell} \bigg)\, ,\\
&\mC^{\ell}_{S_1} = -\frac{v^2}{2\mLQ} (\beta^{b\ell}_L)^\ast \bigg(\beta_L^{b\ell} + \frac{V_{cs}}{V_{cb}} \beta_L^{s\ell} \bigg)\, .
\end{align}
The set of observables that we consider in this category are $R_D, R_{D^\ast}$ and the branching ratio $\mathcal{B}(B_c\to \tau\nu)$. The expressions for $R_D$ and $R_{D^\ast}$ are \cite{Blanke:2018yud} 
\begin{align}
R_D &\approx R_D^{\rm SM}\bigg\{|1+\mC^{\tau}_{V_1}|^2 + 1.54 \re\big[(1+\mC^{\tau}_{V_1})(\mC^{\tau}_{S_1})^\ast\big]\,\nn\\& + 1.09|\mC^{\tau}_{S_1}|^2 \bigg\}\,, \\
R_{D^\ast} &\approx R_{D^\ast}^{\rm SM}\bigg\{|1+\mC^{\tau}_{V_1}|^2 + 0.13 \re\big[(1+\mC^{\tau}_{V_1})(\mC^{\tau}_{S_1})^\ast\big]\,\nn\\& + 0.05|\mC^{\tau}_{S_1}|^2 \bigg\}\,. 
\end{align}
The HFLAV averages that use the most recent measurements of these two observables are $R_D=0.340\pm 0.030$ and $R_{D^\ast} = 0.295\pm 0.013$ \cite{Amhis:2016xyh}. The SM prediction of $R_{D}^{\rm SM}$ and $R_{D^\ast}^{\rm SM}$ are given in the Introduction section. 

\begin{figure}[h!]
	\begin{center}
		\includegraphics[scale=0.25]{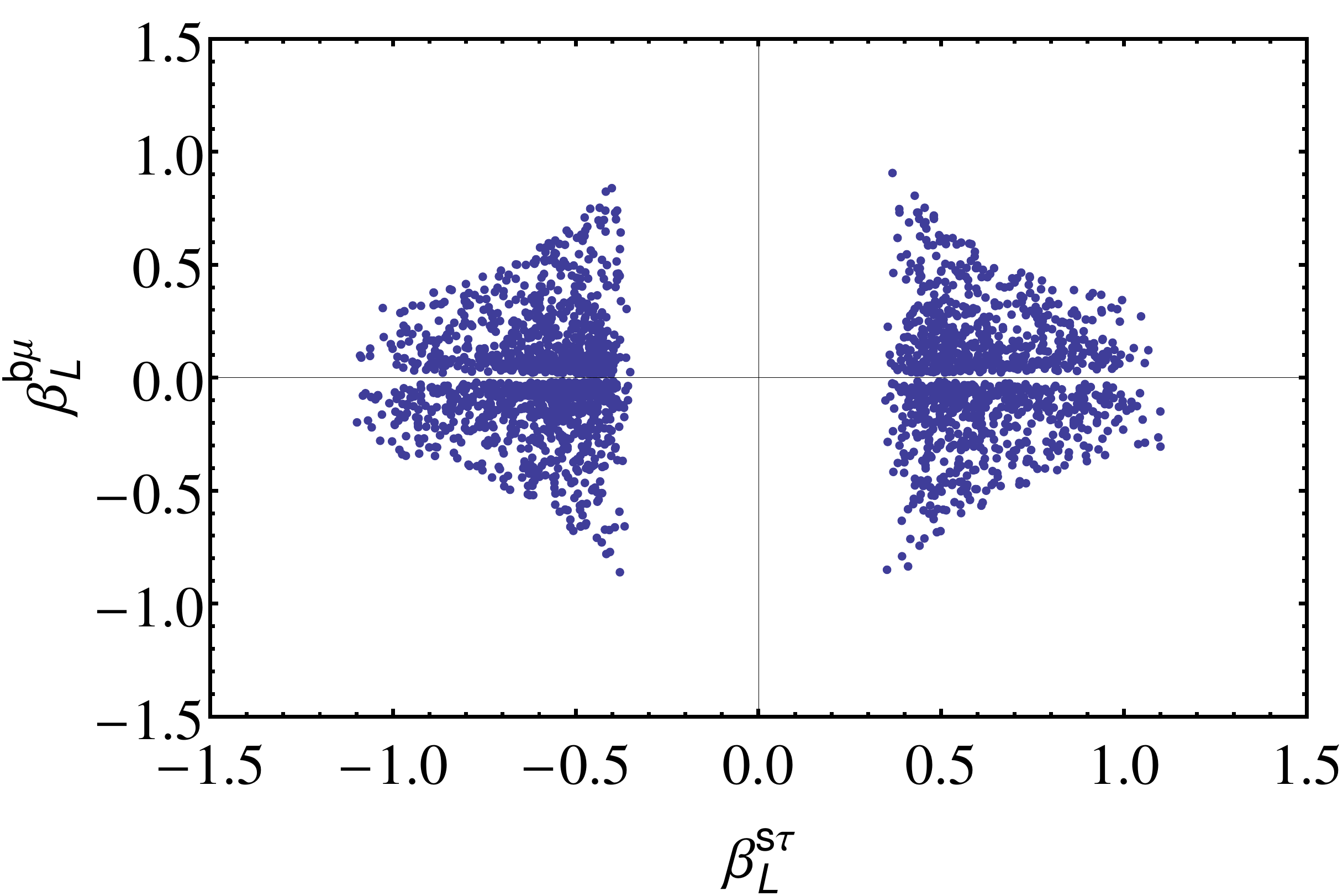}
		\includegraphics[scale=0.25]{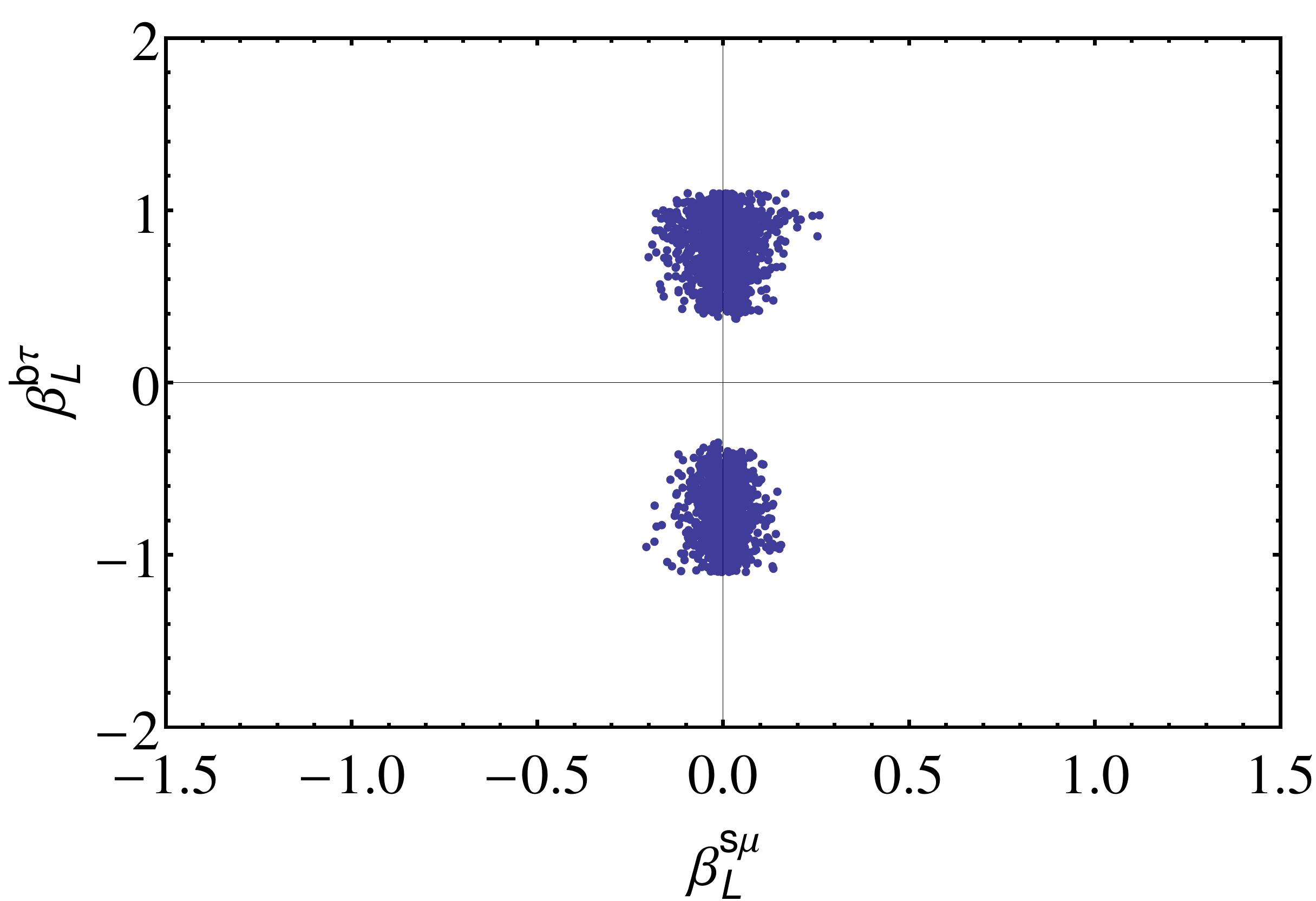}
		\caption{The parameter space (in blue) allowed by low energy observables for the vector leptoquark mass $m_{\rm LQ}=1.5$ TeV. \label{fig:VLQspace}}
	\end{center}
\end{figure}

The $B_c\to\tau\nu$ branching ratio reads  
\begin{align}
&\mathcal{B}(B_c\to \tau\nu) = \frac{\tau_{B_c} m_{B_c} f^2_{B_c}  |V_{cb}|^2 }{16\pi v^4} m_\tau^2 \bigg( 1 - \frac{m_\tau^2}{m^2_{B_c}} \bigg)^2 \bigg| 1+  \,\nn\\& \frac{v^2}{4\mLQ} \bigg( \beta_L^{b\tau}-\frac{2 m^2_{B_s} \beta_R^{b\tau}}{m_\tau(m_b+m_c)}  \bigg)^\ast \bigg( \beta_L^{b\tau} + \frac{V_{cs}}{V_{cb}} \beta_L^{s\tau} \bigg) \bigg|^2\, ,
\end{align}
The most stringent constraint on $B_c\to \tau\nu$ come from LEP data from which Ref.~\cite{Akeroyd:2017mhr} put the limit $\mathcal{B}(B_c\to \tau\nu)\leq 10\%$. Another charged current observable in the $b\to u$ transition that we consider is 
\begin{align}
\mathcal{B}(B\to\tau\nu) &= \mathcal{B}(B\to\tau\nu)_{\rm SM} \bigg|  1+ \frac{v^2}{4\mLQ} \bigg( \beta_L^{b\tau}\, \nn\\&-\frac{2 m^2_{B_s}}{m_\tau(m_b+m_c)} \beta_R^{b\tau} \bigg)^\ast \bigg( \beta_L^{b\tau} + \frac{V_{cs}}{V_{ub}} \beta_L^{s\tau} \bigg) \bigg|^2\, .
\end{align}
\begin{figure}[h!]
	\begin{center}
		\includegraphics[scale=0.25]{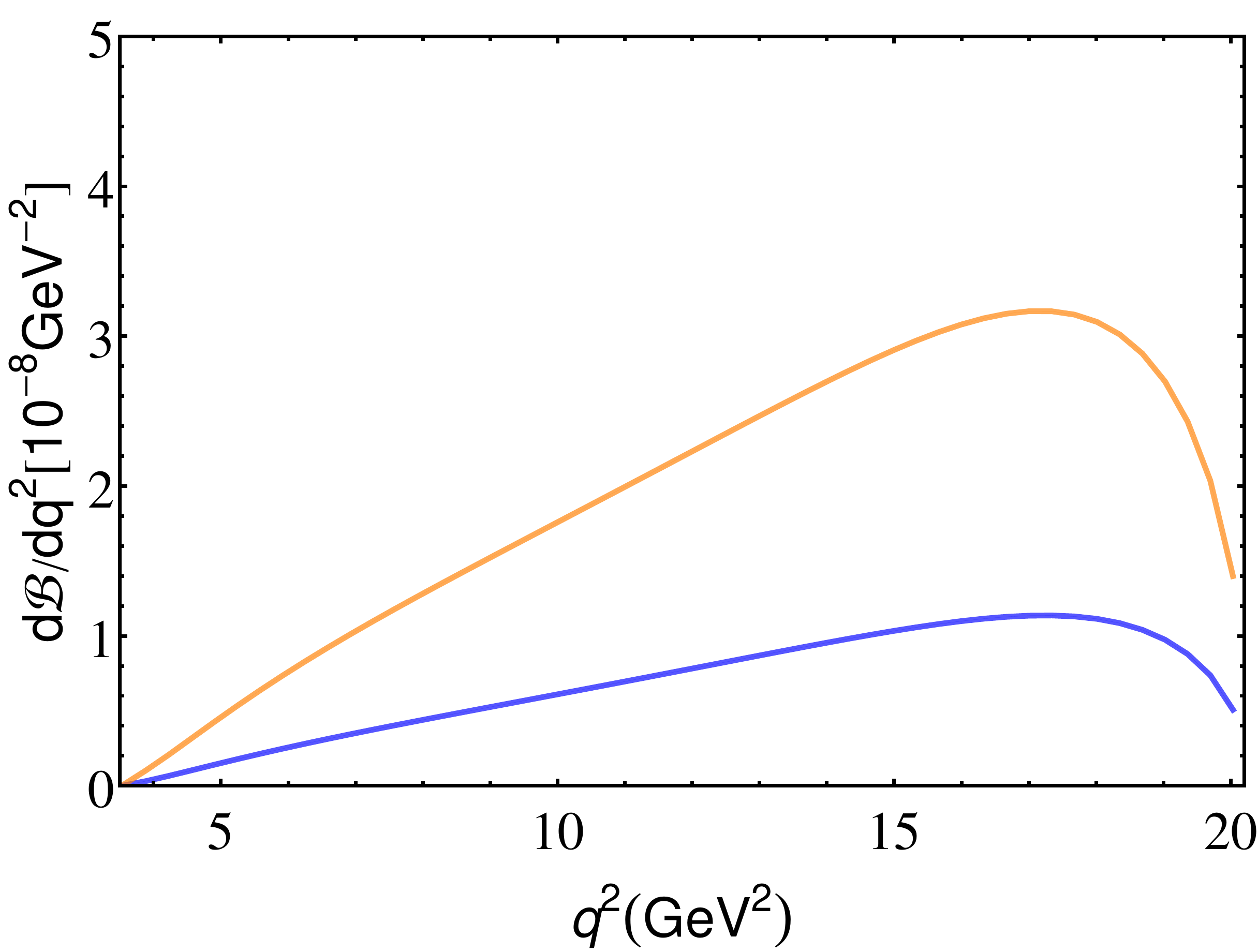}
		\includegraphics[scale=0.265]{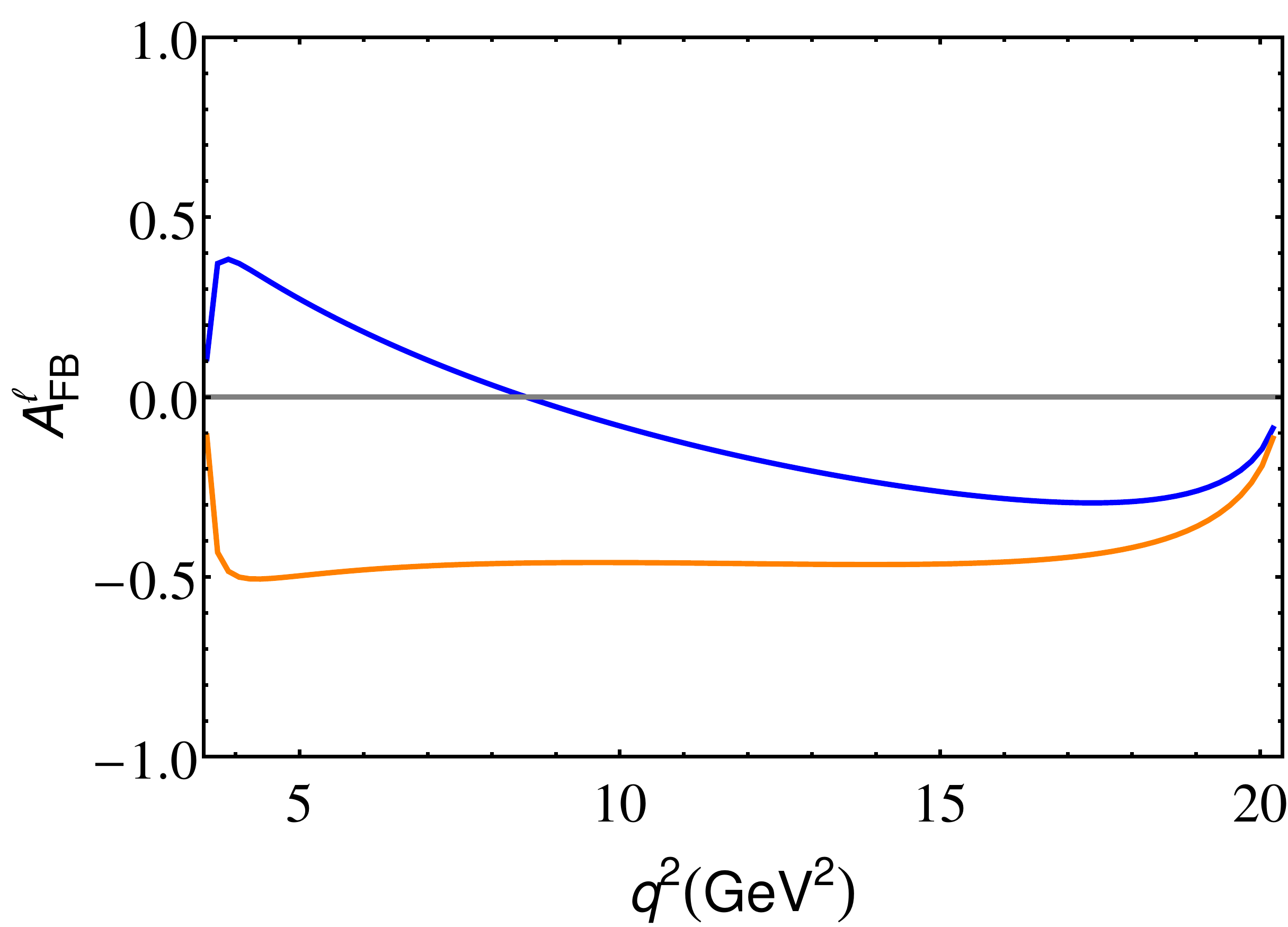}
		\caption{The $q^2$ distribution of the differential branching ratio and the lepton-side forward-backward asymmetry is shown for a set ($\beta^{s\mu}_L=-0.031$, $\beta^{s\tau}_L=0.433$, $\beta^{b\mu}_L=-0.112$,	$\beta^{b\tau}_L=-0.957$,$\beta^{b\tau}_R=-0.128$) of benchmark values of the $U_1$ leptoquark model parameters that are allowed by the low energy observables. The blue and the orange lines correspond to $\Lambda_b\to\Lambda\tau^+\mu^-$ and $\Lambda_b\to\Lambda\mu^+\tau^-$ modes respectively.\label{fig:obs}}
	\end{center}
\end{figure}

According to \cite{Tanabashi:2018oca} $\mathcal{B}(B\to\tau\nu)\leq (1.09 \pm 0.24)\times 10^{-4}$ and $\mathcal{B}(B\to\tau\nu)_{\rm SM} = (0.812 \pm 0.054)\times 10^{4}$ \cite{Bona:2017cxr}. 

We now perform a $\chi^2$ analysis to find the parameter space allowed by the above low energy observables listed in Table \ref{tab:obs}. The $\chi^2$ is defined as
\begin{equation}
	\chi^2 = \sum_i \bigg( \frac{(\mathcal{O}_i^{\rm expt} - \mathcal{O}_i^{\rm th})}{\Delta \mathcal{O}_i^{\rm expt}} \bigg)^2\, ,
\end{equation}
where $\mathcal{O}_i^{\rm expt,(th)}$ are the experimental (theoretical) values of the observables and $\Delta \mathcal{O}_i^{\rm expt}$ are the experimental errors. We minimize the $\chi^2$ and choose a $2\sigma$ region about $\chi^2_{\rm min}$. In this analysis we set mass of the leptoquark $m_{\rm LQ}=1.5$ TeV. In Fig.~\ref{fig:VLQspace} the obtained parameter space is shown. For this parameter space, the $q^2$ distribution of the differential branching ratio and the lepton-side forward-backward asymmetry is shown in Fig.~\ref{fig:obs} for a set of benchmark values of the couplings. The plots are obtained for the central values of the form factors and other inputs. Due to our choice of the flavor structure \eqref{eq:R2flav} the $\Lambda_b\to \mu^+\tau^-$ receives contributions from VA type operators only while the $\Lambda_b\to \tau^+\mu^-$ mode receives contributions from both VA and SP operators. Since in our model $\mC_V=-\mC_A$, in the $\Lambda_b\to\Lambda\mu^+\tau^-$ mode the $A^\ell_{\rm FB}$ is independent of the couplings $\beta_L$ and the forward-backward asymmetry is entirely determined by the form factors and kinematic variables. Interestingly, the $\Lambda_b\to\Lambda\tau^+\mu^-$ mode has a $A^\ell_{\rm FB}$ zero-crossing which is absent in the $\Lambda_b\to\Lambda\mu^+\tau^-$ mode. For the obtained parameter space we also calculate the maximum and the minimum values of the branching ratio and $A^\ell_{\rm FB}$ integrated over the entire $q^2$ phase space,
\begin{align}
&\langle \mathcal{B}(\Lambda_b\to\Lambda\tau^+\mu^-)\rangle = [1.55\times 10^{-9} ,7.83\times 10^{-6}]\, ,\\
&\langle\mathcal{B}(\Lambda_b\to\Lambda\mu^+\tau^-)\rangle = [5.01\times 10^{-9}, 1.78\times 10^{-5}]\, ,\\
%\end{align}
%%
%and
%%
%\begin{align}
&\langle A^\ell_{\rm FB}(\Lambda_b\to\Lambda\tau^+\mu^-)\rangle = [-0.2504,-0.003]\, ,\\
&\langle A^\ell_{\rm FB}(\Lambda_b\to\Lambda\mu^+\tau^-)\rangle = -0.4040\, .
\end{align}
The large branching ratios of the order $\mathcal{O}(10^{-5},10^{-6})$ are induced by large ranges of $\beta^{q\tau}$ allowed by the current data. Such large ranges arise due to poor experimental bounds on modes such as $B_s\to\tau^+\tau^-, B^+\to K\tau^+\tau^-$. These branching ratios are accessible in the LHCb.

\begin{table}[ht]
\centering
\renewcommand{\arraystretch}{1.2} 
\begin{tabular}{|c |c |c|c }
\hline
Observable & Experiment & SM  \\
\hline 
$\Delta \mathcal{C}_{V}^{\mu \mu}=-\Delta \mathcal{C}_{A}^{\mu \mu}$ &  $[-0.59,-0.40]$ \cite{Alguero:2019ptt} \cite{Aebischer:2019mlg}  & --   \\
$\Delta \mathcal{C}_{V}$ &  $-0.73\pm 0.23$ \cite{Alguero:2019ptt} \cite{Aebischer:2019mlg} & $-$  \\
$\mathcal{B}(B_{s} \to \tau^{+} \tau^{-})$ & $0.0(3.4) \cdot 10^{-3}$ \cite{Aaij:2017xqt}  &$7.73(49)\cdot 10^{-7}$ \cite{Bobeth:2013uxa} \\
$\mathcal{B}(B^+ \to K^+ \tau^{+} \tau^{-})$ & $1.36(0.71) \cdot 10^{-3}$ \cite{TheBaBar:2016xwe}  &$1.44(0.28)\cdot 10^{-7}$ \\
$\mathcal{B}(B^{+} \to K^{+} \tau^{+} \mu^{-}) $ & $2.8 \cdot 10^{-5}$ \cite{Lees:2012zz}  & $-$  \\
$\mathcal{B}(B^{+} \to K^{+} \mu^{+} \tau^{-}) $ & $4.5\cdot 10^{-5}$ \cite{Lees:2012zz}  & $-$  \\
$\mathcal{B}( \tau \to \mu \phi)$ & $0.0(5.1) \cdot 10^{-8}$ \cite{Miyazaki:2011xe}& $-$   \\
$R_{D} $& 0.340(30) \cite{Amhis:2016xyh}  &  $0.299(3)$ \cite{Bigi:2016mdz}\\
$R_{D^\ast} $& 0.295(13) \cite{Amhis:2016xyh}  &  $0.258(5)$ \cite{Jaiswal:2017rve}\\

$\mathcal{B}( B \to \tau \nu)$ &$1.09(24)\cdot 10^{-4}$ \cite{Tanabashi:2018oca}  &$0.812(54) \cdot 10^{-4}$ \cite{Bona:2017cxr}  \\
$\mathcal{B}(\tau \to \mu \gamma) $ & $0.0(3.0) \cdot 10^{-8}$ \cite{Amhis:2016xyh}  & $-$ \\
$\mathcal{B}(\tau \to \mu \phi) $ & $0.0(5.1) \cdot 10^{-8}$ \cite{Miyazaki:2011xe}  & $-$ \\
\hline
\end{tabular}
\caption{List of observables included in the fit. \label{tab:obs}}
\end{table}

%%%%%%%%%%%%%%%%%%%%%%%%%%%%%%%%%%%%%%%%%%%%%%%%%%%%%%%%%%%%%%%
%%%%%%%%%%%%%%%%%%%%%%%%%%%%%%%%%%%%%%%%%%%%%%%%%%%%%%%%%%%%%%%
\section{Summary \label{sec:summary}}
Lepton flavor violating decays are strictly forbidden in the Standard Model and therefore any observation is a smoking gun signal of physics beyond the Standard Model. In recent years a number of lepton flavor universality violating decays has been observed albeit of low statistical significance. Many physics beyond the Standard Models that has been constructed to explain the origin of flavor universality violating couplings can also give rise to flavor violating decays. Motivated by these results, in this paper we have explored lepton flavor violating $b\to s\ell_1^+\ell_2^-$ transition in $\Lambda_b\to \Lambda\ell_1^+\ell_2^-$ decay. In this paper we have presented a double differential distribution of the decay in terms of dilepton invariant mass squared $q^2$ and lepton angle $\theta_\ell$. From this distribution we have obtained the differential branching ratio and the lepton-side forward-backward asymmetry. We have studied these two observables in the vector leptoquark model $U_1\equiv (\textbf{3,1})_{2/3}$. The parameter space of the model has been constrained by low energy observables. Our predicted range of the branching ratio in the vector leptoquark model may be accessible by the LHCb.

%%%%%%%%%%%%%%%%%%%%%%%%%%%%%%%%%%%%%%%%%%%%%%%%%%%%%%%%%%%%%%%
%%%%%%%%%%%%%%%%%%%%%%%%%%%%%%%%%%%%%%%%%%%%%%%%%%%%%%%%%%%%%%%
\section*{Acknowledgements}
The author is supported by the DST, Govt. of India under INSPIRE Faculty Award.
%%%%%%%%%%%%%%%%%%%%%%%%%%%%%%%%%%%%%%%%%%%%%%%%%%%%%%%%%%%%%%%
%%%%%%%%%%%%%%%%%%%%%%%%%%%%%%%%%%%%%%%%%%%%%%%%%%%%%%%%%%%%%%%

\appendix
%%%%%%%%%%%%%%%%%%%%%%%%%%%%%%%%%%%%%%%%%%%%%%%%%%%%%%%%%%%%%%%
%%%%%%%%%%%%%%%%%%%%%%%%%%%%%%%%%%%%%%%%%%%%%%%%%%%%%%%%%%%%%%%
\section{Transversity amplitudes \label{sec:TAs2}}
Corresponding to the effective Hamiltonian \eqref{eq:Heff1} the expressions of the transversity amplitudes read \cite{Das:2018iap}
\begin{eqnarray}
A^{L,(R)}_{\perp_1} &=& -\sqrt{2}N \bigg( f^V_\perp \sqrt{2s_-} \mC^{L,(R)}_{\rm VA+} \bigg)\, ,\\
A^{L,(R)}_{\|_1} &=& \sqrt{2}N \bigg( f^A_\perp \sqrt{2s_+} \mC^{L,(R)}_{\rm VA-}  \bigg)\, ,\\
A^{L,(R)}_{\perp_0} &=& \sqrt{2}N \bigg( f^V_0 (\mLb + \mL) \sqrt{\frac{s_-}{q^2}} \mC^{L,(R)}_{\rm VA+}  \bigg)\, ,\\
A^{L,(R)}_{\|_0} &=& -\sqrt{2}N \bigg( f^A_0 (\mLb - \mL) \sqrt{\frac{s_+}{q^2}} \mC^{L,(R)}_{\rm VA-}  \bigg)\, ,\\
A_{\perp t} &=& -2\sqrt{2}N f^V_t (\mLb - \mL) \sqrt{\frac{s_+}{q^2}} (\mC_{A} + \mC_A^\prime)\, ,\\
A_{\| t} &=& 2\sqrt{2}N f^A_t (\mLb + \mL) \sqrt{\frac{s_-}{q^2}} (\mC_{A} - \mC_A^\prime) \, .
\end{eqnarray}
Here the normalization constant $N(q^2)$ is given by 
\begin{align}
&N(q^2) =  \frac{V_{tb}V_{ts}^\ast \alpha_e}{\sqrt{2}v^2} \sqrt{ \tau_{\Lambda_b} \frac{q^2 \sqrt{\lambda(\mmLb,\mmL,q^2)} }{2^{15} m^3_{\Lambda_b} \pi^5}	\beta_\ell \beta_\ell^\prime}\, ,\nn\\
&\beta_\ell = \sqrt{1 - \frac{(m_1+m_2)^2}{q^2}}\,,\quad \beta_\ell^\prime = \sqrt{1 - \frac{(m_1-m_2)^2}{q^2}}\, ,
\end{align}
where $\lambda(a,b,c)=a^2 + b^2 + c^2 - 2(ab + bc + ca)$ and the Wilson coefficients are
\begin{align}
& \mC_{\rm VA,+}^{L(R)} = (\mC_V\mp \mC_A)+(\mC_V^\prime \mp \mC_A^\prime)\, ,\\
& \mC_{\rm VA,-}^{L(R)} = (\mC_V\mp \mC_A)-(\mC_V^\prime \mp \mC_A^\prime)\, .
\end{align}
The transversity amplitudes corresponding to the SP operators are \cite{Das:2018iap}
\begin{eqnarray}
A_{\rm S\perp} &=& 2\sqrt{2}N f^V_t \frac{\mLb - \mL}{m_b} \sqrt{s_+} (\mC_S + \mC_S^\prime)\, ,\\
A_{\rm S\|} &=& -2\sqrt{2}N f^A_t \frac{\mLb + \mL}{m_b} \sqrt{s_-} (\mC_S - \mC_S^\prime)\, ,\\
A_{\rm P\perp} &=& -2\sqrt{2}N f^V_t \frac{\mLb - \mL}{m_b} \sqrt{s_+} (\mC_P + \mC_P^\prime)\, ,\\
A_{\rm P\|} &=& 2\sqrt{2}N f^A_t \frac{\mLb + \mL}{m_b} \sqrt{s_-} (\mC_P - \mC_P^\prime)\, .
\end{eqnarray}

%%%%%%%%%%%%%%%%%%%%%%%%%%%%%%%%%%%%%%%%%%%%%%%%%%%%%%%%%%%%%%%
%%%%%%%%%%%%%%%%%%%%%%%%%%%%%%%%%%%%%%%%%%%%%%%%%%%%%%%%%%%%%%%
\section{Spinors in dilepton rest frame \label{sec:llRF}}
We assume that the lepton $\ell_2^-$ is negatively charged and has four-momentum is $q_2^\mu=(E_1,\vec{q})$, while $\ell_1^+$ is positively charged and has four-momentum $q_1^\mu=(E_1,-\vec{q})$ 
\begin{align}
& q_1^\mu \Big|_{2\ell} = (E_2, -|q_{2\ell}|\sin\theta_\ell, 0, -|q_{2\ell}|\cos\theta_\ell)\, ,\\
&q_2^\mu \Big|_{2\ell} = (E_1, |q_{2\ell}|\sin\theta_\ell, 0, |q_{2\ell}|\cos\theta_\ell)\, ,
\end{align}
with 
\begin{eqnarray}
|q_{2\ell}| &=& \frac{\lambda^{1/2}(q^2,m_1^2,m_2^2)}{2\sqrt{q^2}}\, ,\quad\quad E_1 = \frac{q^2+m_1^2-m_2^2}{2\sqrt{q^2}}\,,\nn\\  E_2 &=& \frac{q^2+m_2^2-m_1^2}{2\sqrt{q^2}}\, .
\end{eqnarray}
The explicit expressions of the lepton helicity amplitudes require us to calculate 
\begin{equation}
\bar{u}_{\ell_2} (1\mp\gamma_5) v_{\ell_1}\, ,\quad \bar{\epsilon}^\mu(\lambda) \bar{u}_{\ell_2} \gamma_\mu (1\mp\gamma_5) v_{\ell_1}\, .
\end{equation}
Following \cite{Haber:1994pe} the explicit expressions of the spinor for the lepton $\ell_2^-$ is 
\begin{align}
& u_{\ell_2}(\lambda) = 
\begin{pmatrix}
\sqrt{E_\ell+m_\ell} \chi^u_\lambda  \\ 2 \lambda \sqrt{E_\ell-m_\ell} \chi^u_\lambda
\end{pmatrix}\, ,
\quad \chi^u_{+\frac{1}{2}} = \begin{pmatrix} \cos\frac{\theta_\ell}{2} \\ \sin\frac{\theta_\ell}{2} \end{pmatrix}\, \nn\\
&\chi^u_{-\frac{1}{2}} = \begin{pmatrix} -\sin\frac{\theta_\ell}{2} \\ \cos\frac{\theta_\ell}{2} \end{pmatrix}\, .
\end{align}
For the lepton $\ell_1^+$ which is moving in the opposite direction to $\ell_2$, the two component spinor $\chi^v$ looks like
\begin{equation}
\chi^v_{-\lambda} = \xi_\lambda \chi^u_{\lambda}\, ,\quad \xi_\lambda = 2\lambda e^{-2i\lambda\phi}\, .
\end{equation}
Hence we have 
\begin{align}
& v_{\ell_1}(\lambda) = 
\begin{pmatrix}
\sqrt{E_\ell-m_\ell} \chi^v_{-\lambda}  \\ -2 \lambda \sqrt{E_\ell+m_\ell} \chi^v_{-\lambda}
\end{pmatrix}\, ,
\quad \chi^v_{+\frac{1}{2}} = \begin{pmatrix} \sin\frac{\theta_\ell}{2} \\ -\cos\frac{\theta_\ell}{2} \end{pmatrix}\, \nn\\
&\chi^v_{-\frac{1}{2}} = \begin{pmatrix} \cos\frac{\theta_\ell}{2} \\ \sin\frac{\theta_\ell}{2} \end{pmatrix}\, .
\end{align}

With these choices of lepton spinors we get the following expressions of the lepton helicity amplitudes $L^{\lambda_2,\lambda_1}_{L(R)}$ and $ L^{\lambda_2,\lambda_1}_{L(R),\lambda}$
\begin{align}
&L_L^{\plpl} = \sqrt{q^2}(\beta_\ell^\prime+\beta_\ell)\, ,\quad L_L^{\plmi} = 0\, ,\quad L_L^{\mipl} = 0\, \nn\\& L_L^{\mimi} = \sqrt{q^2}(\beta_\ell^\prime-\beta_\ell)\, ,\\
& L_R^{\plpl} = -\sqrt{q^2}(\beta_\ell^\prime - \beta_\ell)\, ,\quad L_R^{\plmi} = 0\, ,\quad L_R^{\mipl} = 0\, ,\nn\\ &L_R^{\mimi} = -\sqrt{q^2}(\beta_\ell^\prime + \beta_\ell)\, ,\\
& L_{L,+1}^{\plpl} = \frac{1}{\sqrt{2}}\big[m_1(\beta_\ell^\prime +\beta_\ell) + m_2(\beta_\ell^\prime - \beta_\ell) \big]\sl\, ,\\
& L_{L,+1}^{\plmi} = -\sqrt{\frac{q^2}{2}} (\blp-\bl) (1-\cl)\, ,\nn\\ &L_{L,+1}^{\mipl} = \sqrt{\frac{q^2}{2}}(\blp+\bl)(1+\cl)\, ,\\
& L_{L,+1}^{\mimi} = -\frac{1}{\sqrt{2}}\big[m_1(\blp - \bl) + m_2(\blp + \bl) \big]\, ,\\
& L_{R,+1}^{\plpl} = \frac{1}{\sqrt{2}}\bigg[ m_1(\blp-\bl) + m_2(\blp+\bl) \bigg]\sl\, ,\\
& L_{R,+1}^{\plmi} = -\sqrt{\frac{q^2}{2}} (\blp+\bl)(1-\cl)\, ,\nn\\ &L_{R,+1}^{\mipl} = \sqrt{\frac{q^2}{2}} (\blp-\bl) (1+\cl)\, ,\\
& L_{R,+1}^{\mimi} = -\frac{1}{\sqrt{2}} \big[ m_1(\blp+\bl) + m_2(\blp-\bl) \bigg]\sl\, ,\\
& L_{L,-1}^{\plpl} = -\frac{1}{\sqrt{2}} \big[m_1(\blp+\bl)+m_2(\blp-\bl) \big]\, ,\\
& L_{L,-1}^{\plmi} = -\sqrt{\frac{q^2}{2}} (\blp-\bl) (1+\cl)\, ,\nn\\ &L_{L,-1}^{\mipl} = \sqrt{\frac{q^2}{2}} (\blp+\bl) (1-\cl)\, ,\\
& L_{L,-1}^{\mimi} = \frac{1}{\sqrt{2}} \big[m_1(\blp-\bl) + m_2(\blp+\bl) \big]\sl\, ,\\
& L_{R,-1}^{\plpl} = -\frac{1}{\sqrt{2}} \big[m_1(\blp-\bl) + m_2(\blp+\bl) \big]\sl\, ,\\
& L_{R,-1}^{\plmi} = \sqrt{\frac{q^2}{2}}(\blp+\bl)(1+\cl)\, ,\nn\\ &L_{R,-1}^{\mipl} = \sqrt{\frac{q^2}{2}}(\blp-\bl)(1-\cl)\, ,\\
& L_{R,-1}^{\mimi} = \frac{1}{\sqrt{2}}\big[m_1(\blp+\bl) + m_2(\blp-\bl) \big]\sl\, ,\\
& L_{L,0}^{\plpl} = -\big[m_1(\blp+\bl)-m_2(\blp-\bl) \big]\cl\, ,\\
& L_{L,0}^{\plmi} = \sqrt{q^2}(\blp-\bl)\cl\, ,\nn\\ &L_{L,0}^{\mipl}=\sqrt{q^2}(\blp+\bl)\cl\, ,\\
& L_{L,0}^{\mimi} = \big[m_1(\blp-\bl) + m_2(\blp+\bl) \big]\cl\, ,\\
& L_{R,0}^{\plpl} = -\big[m_1(\blp-\bl) + m_2(\blp+\bl) \big]\cl\, ,\\
& L_{R,0}^{\plmi} = \sqrt{q^2}(\blp+\bl)\cl\,,\nn\\ &L_{R,0}^{\mipl} =\sqrt{q^2}(\blp-\bl)\sl\,,\\
& L_{R,0}^{\mimi} = \big[m_1(\blp+\bl) + m_2(\blp-\bl) \big]\cl\, ,\\
& L_{L,0}^{\plpl} = \big[m_1(\blp+\bl) + m_2(\blp-\bl) \big]\, ,\nn\\ &L_{L,0}^{\plmi} = L_{L,0}^{\mipl} = 0\, ,\\
& L_{L,0}^{\mimi} = \big[m_1(\blp-\bl) + m_2(\blp+\bl) \big]\, ,\\
& L_{R,0}^{\plpl} = -\big[m_1(\blp-\bl) + m_2(\blp+\bl) \big]\, ,\nn\\&L_{R,0}^{\plmi} = L_{R,0}^{\mipl} = 0\, ,\\
& L_{R,0}^{\mimi} = -\big[m_1(\blp+\bl) + m_2(\blp-\bl) \big]\, .
\end{align}

\end{document}